\documentclass[12pt,twoside]{article}
\usepackage[dvips]{graphicx}
\usepackage{float}
\usepackage{latexsym}
\usepackage{amsmath,mathrsfs,bm,amssymb,color}
\newtheorem{theorem}{Theorem}[section]
\newtheorem{proposition}[theorem]{Proposition}
\newtheorem{lemma}[theorem]{Lemma}
\newtheorem{corollary}[theorem]{Corollary}
\newtheorem{definition}[theorem]{Definition}
\newtheorem{example}[theorem]{Example}
\newtheorem{remark}[theorem]{Remark}

\newtheorem{assumption}[theorem]{Assumption}

\newcommand{\bd}[1]{\begin{definition}\label{#1}}
\newcommand{\ed}{\end{definition}}
\newcommand{\la}{\lambda}
\newcommand{\bl}[1]{\begin{lemma}\label{#1}}
\newcommand{\el}{\end{lemma}}
\newcommand{\bc}[1]{\begin{corollary}\label{#1}}
\newcommand{\ec}{\end{corollary}}
\newcommand{\bt}[1]{\begin{theorem}\label{#1}}
\newcommand{\et}{\end{theorem}}
\newcommand{\bp}[1]{\begin{proposition}\label{#1}}
\newcommand{\ep}{\end{proposition}}
\newcommand{\br}[1]{\begin{remark}\label{#1}}
\newcommand{\er}{\end{remark}}
\newcommand{\eq}[1]{\begin{equation}\label{#1}}
\newcommand{\en}{\end{equation}}
\newcommand{\eqn}{\begin{eqnarray*}}
\newcommand{\enn}{\end{eqnarray*}}
\newcommand{\eqnn}{\begin{eqnarray}}
\newcommand{\ennn}{\end{eqnarray}}
\newcommand{\proof}{{\noindent {\sc Proof}: \,}}
\newcommand{\qed}{\hfill {\bf qed}\par\medskip}

\newcommand{\CC}{{{\mathbb C}}}
\newcommand{\RR}{{\mathbb R}}

\newcommand{\kak}[1]{(\ref{#1})}

\newcommand{\ov}[1]{\overline{#1}}

\newcommand{\lk}{\left(}
\newcommand{\rk}{\right)}
\newcommand{\lkk}{\left\{}
\newcommand{\rkk}{\right\}}

\renewcommand{\d}{\displaystyle}

\newcommand{\hhh}{\mathscr{H}}
\newcommand{\ms}[1]{\mathscr{#1}}
\newcommand{\dt}{{\dm(T)}}

\newcommand{\half}{\frac{1}{2}}
\newcommand{\han}{{1/2}}

\newcommand{\dm}{{\rm D}}

\newcommand{\g}{g'(H)^{-1}}

\newcommand{\gbb}{g'(H)^{-1}}
\newcommand{\e}{e^{-itg(H)}}

\makeatletter
\@addtoreset{equation}{section}
\makeatother
\def\theequation{\arabic{section}.\arabic{equation}}

\begin{document}

\makeatletter \@addtoreset{equation}{section} \makeatother
\def\theequation{\arabic{section}.\arabic{equation}}
\title
{\sc Strong time operators associated with
generalized Hamiltonians}
\author{\sc Fumio Hiroshima\thanks{F.H. thanks for  Grant-in-Aid for Science Research (B) 20340032 from JSPS for financial support.
}
\thanks{Graduate School of Mathematics,
 Kyushu University
 812-8581, Fukuoka, Japan},
  Sotaro Kuribayashi\thanks{Graduate School of Mathematics,
 Kyushu University
 812-8581, Fukuoka, Japan}\\
and Yasumichi Matsuzawa\thanks{Department of Mathematics, Hokkaido University, Sapporo, 060-0810, Japan}
}

\date{}
\pagestyle{myheadings}
\markboth{Time operators}
{Time operators}

\setlength{\baselineskip}{18pt}
\maketitle

\begin{abstract}
Let the pair of operators, $(H,T)$, satisfy
the weak Weyl relation:
$$Te^{-itH}=e^{-itH}(T+t),$$
where
$H$ is self-adjoint and $T$ is closed symmetric.
Suppose that $g$ is a real-valued Lebesgue measurable
function on $\RR$ such that
$g\in C^2(\RR\setminus K)$
      for  some closed subset $K\subset\RR$
      with Lebesgue measure zero.
Then we can construct a closed symmetric operator $D$ such that $(g(H),D)$ also obeys the weak Weyl relation.
\end{abstract}

\section{Weak Weyl relation and strong time operators}
\subsection{Introduction}
The energy of a quantum system can be realized
as a self-adjoint operator on some Hilbert space,
whereas time $t$ is treated as a parameter, and not intuitively as an operator.
So, since the foundation of quantum mechanics,
the energy-time uncertainty relation has had
a different basis from that underlying
the position-momentum uncertainty relation.

Let $Q$ be the multiplication operator defined by $(Qf)(x)=xf(x)$ with maximal domain
$\d \dm(Q)=\{f\in L^2(\RR)|\int |x|^2 |f(x)|^2 dx<\infty\}$ and let
$P=-id/dx$ be the weak derivative
with domain $H^1(\RR)$.
In quantum mechanics,
the position operator $Q$
and the momentum operator $P$ in $L^2(\RR)$
obey the Weyl relation:
$\d e^{-isP}e^{-itQ}=e^{ist}e^{-itQ}e^{-isP}$ for $
s,t\in\RR$.
From this we can derive
the so-called weak Weyl relation:
\eq{1111}
Q e^{-it P}=e^{-itP} (Q+t),\quad t\in\RR,
\en
and moreover
the canonical commutation relation
$[P,Q]=-iI$
also holds.
%We have to be careful, however, that converse is not true.
The strong time operator $T$ is defined as an operator satisfying \kak{1111} with
$Q$ and $P$ replaced by $T$ and the Hamiltonian $H$ of the quantum system
under consideration, respectively.

More precisely,
we explain the weak Weyl relation \kak{1111} as follows.
Let $\hhh$ be a Hilbert space over the complex field $\CC$.
We denote by
$\dm(L)$ the domain of an operator $L$.
\begin{definition}
\label{im1}
We say that the pair $(H,T)$ consisting of a self-adjoint operator $H$ and a symmetric operator $T$ on $\hhh$
  obeys the weak Weyl relation if and only if, for all $t\in\RR$,
   \begin{description}
  \item[(1)]
   $e^{-itH}\dm(T)\subset \dm(T)$;
  \item[(2)] $Te^{-itH}\Phi=e^{-itH}(T+t)\Phi$ for all $\Phi\in \dm(T)$.
\end{description}
\end{definition}
Here $T$ is referred    to as a strong time operator associated with $H$ and
we denote it
by $T_H$ for $T$.
Note that a strong time operator is not unique.
Although from the weak Weyl relation it follows that
$[H,T_H]=-iI$, the converse is
not true; a pair $(A,B)$ satisfying $[A,B]=-iI$ does not necessarily obey
the Weyl relation or
the weak Weyl relation.
If  strong time operator $T_H$ is self-adjoint, then
it is known that
\eq{st}
e^{-isT_H}e^{-itH}=e^{-ist} e^{-itH}e^{-isT_H}
\en
holds.
In particular when Hilbert space $\hhh$ is separable, by the von Neumann uniqueness theorem
 the Weyl relation \kak{st} implies that $H$ and $T_H$ are unitarily equivalent to
 $\oplus^n P$ and $\oplus^n Q$ with some $n$, respectively.
This asserts that
any strong time operators
     associated with a semibounded $H$
on a separable Hilbert space  are symmetric non-self-adjoint. These facts may implicitly suggest that
 strong time operators are not "observable".

A time operator but not necessarily strong
 associated with a self-adjoint operator $H$ is defined as
an operator $T$ for which
$[H,T]=-iI$.
%A strict definition of weak time operators is given in \cite{am081}.
As was mentioned above,
 although a strong time operator is automatically
a time operator,
the converse is not true.
It is remarkable that when the pair $(H,T)$ obeys
the weak Weyl relation, $H$ has purely absolutely continuous spectrum.
For example there is  no strong time operator associated with
the harmonic oscillator $\half(P^2+\omega^2 Q^2)$,
 whereas its time operator is
formally given by
$$\frac{1}{2\omega}
(\arctan (\omega P^{-1}Q)+\arctan (\omega QP^{-1})).$$
See e.g. \cite{a08b, am081, gal02,gal04, llh90,dor84,ros69}.

The concept of time operators was derived in the framework
for the energy-time uncertainty relation in \cite{dk94}.
%, where  weak time operators are considered.
See also e.g. \cite{fuj80,fwy80,gys81-1,gys81-2}. A strong  connection
with the decay of survival probability  was
pointed out by \cite{m01}, where the weak Weyl relation was introduced and then strong time operators were discussed.
Moreover it was drastically generalized in \cite{a05} and some uniqueness
theorems are established in \cite{a08}.

This paper is inspired by
\cite[Section VII]{m01} and
\cite{am08}.
In particular
Arai and Matsuzawa \cite{am08} developed machinery for reconstructing a pair of operators obeying the weak Weyl relation from a given pair $(H,T_H)$; in particular, they constructed a strong time operator associated with $\log|H|$.
The main result of the paper is an extension of this work and 
we derive a strong time operator
associated with  general Hamiltonian $g(H)$ with a real-valued function $g$.

\subsection{Description of the  main results}
By \kak{1111} a strong time operator $T_P$
associated with $P$ is given by
\eq{112}
T_P=Q.
\en
 For the self-adjoint operator  $(\han)P^2$ in $L^2(\RR)$, it is established that
\eq{111}
T_{(\han)P^2}=\half (P^{-1}Q+QP^{-1})
\en
is an associated strong time operator
referred to as the Aharonov-Bohm operator.
Comparing \kak{112} with \kak{111} we arrive at
\eq{113}
T_{(\han)P^2}=\half \lk f'(P)^{-1} T_P+T_P f'(P)^{-1} \rk,
\en
where $f(\lambda)=(\han)\lambda^2$.
We wish to extend formula \kak{113}
for more general $f$'s
and for any  $(H,T_H)$.

More precisely let $g$ be some Borel measurable function from $\RR$ to $\RR$.
We want to construct a map $\ms T(g)$
such that
$\ms T (g)T_{H}=T_{g(H)}$ and to show that
$$T_{g(H)}=\half(g'(H)^{-1} T_H+T_H g'(H)^{-1}).$$

We denote the set of $n$ times continuously differentiable functions on $\Omega\subset \RR$ with compact support by $C_0^n(\Omega)$.
\bp{im3}
Assume that $(H,T)$ satisfies the weak Weyl relation.
Then
\begin{itemize}
\item[(1)] $H$ has purely absolutely continuous spectrum. In particular $H$ has no point spectrum;
    \item[(2)]
    $(H,\ov{T})$ also satisfies the weak Weyl relation.
    \end{itemize}
\ep
\proof
(1) Refer to see \cite{a05}.
(2) It can be proven by a simple limiting argument. \qed
         Throughout, we suppose that the following assumptions hold.
\begin{assumption}
\label{a1}
      $(H,T)$ obeys the weak Weyl relation and $T$ is a {\it closed} symmetric operator.
  \end{assumption}

\begin{assumption}
\label{a2}
   Let $g:\RR\rightarrow\RR$ be a Lebesgue measurable function such that
\begin{itemize}
\item[(1)]
$g\in C^2(\RR\setminus K)$
for some closed subset $K\subset \RR$ with Lebesgue measure zero;
\item[(2)] the Lebesgue measure of $\lkk
\la\in \RR\setminus K|g'(\la)=0\rkk$ is zero.
\end{itemize}
   \end{assumption}
We fix $(H,T)$, $K\subset \RR$ and $g\in C^2(\RR\setminus K)$
satisfying Assumptions \ref{a1} and \ref{a2}
in what follows.
For a Lebesgue measurable function $f$, $f(H)$ is defined by
$$f(H)=\int_{{\rm Spec}(H)} f(\la) dE_\la^H$$
 for the spectral resolution $E_\la^H$ of $H$.
Let $Z$ be the set of singular points of $g'^{-1}$:
$$Z=\{\la\in\RR\setminus K|g'(\la)=0\}
\cup K,$$
which is closed and has Lebesgue measure zero.

Now we will define a useful subspace $X_n^\ms D$.
\bd{im5}
Let $\ms D\subset \ms H$ be a dense subspace.
The subspace
 $X_n^{\ms D}$, $0\leq n\leq \infty$,
 in $\hhh$ is defined by
\eq{3}
X_n^{\ms D}=\mbox{\rm linear hull of }\{\rho(H)\phi|\rho\in C_0^n(\RR\setminus Z),\phi\in \ms D\},
\en
where $C_0^0=C_0$.
\ed
\bl{im2}
$X_n^\ms D$ is dense in $\ms H$.
\el
\proof
Let $(f,\Phi)=0$ for all $\Phi\in X_n^{\ms D}$.
Then $(\rho(H)^\ast f,\phi)=0$ for all $\phi\in \ms D$ and $\rho\in C_0^n(\RR\setminus Z)$,  which implies that $f\in E_{Z}^H\ms H$, where $E^H_{\cdot}$ denotes the spectral resolution of $H$.
Since $H$ has purely absolutely continuous spectrum and the Lebesgue measure of $Z$ is zero, $f=0$ is concluded. Hence $X_n^\ms D$ is dense.
\qed

The next proposition is fundamental.
\bp{l1}
{\rm \cite{a05}}
Let $f\in C^1(\RR)$ and let both $f$ and $f'$ be bounded.
Then $f(H)\dm(T)\subset \dm(T)$ and
\eq{arai}
Tf(H)\phi=f (H)T\phi+i f'(H)\phi,\quad \phi\in \dm(T).
\en
\ep
\proof
First suppose that $f\in C_0^\infty(\RR)$. Let $\check f$ denote the inverse Fourier transform of $f$.
Then
for $\psi\in \dm(T)$,
\begin{eqnarray*}
(T\psi, f(H)\phi)
&=&
(2\pi)^{-\han}
\int_{\RR} (T\psi,
e^{-i\la H}\phi)\check{f}(\la)d\la\\
&=&
(2\pi)^{-\han}\int_{\RR}\check{f}(\la)(\psi, e^{-i\la H}(T+\la)\phi) d\la=(\psi, (f(H)T+if '(H))\phi).
\end{eqnarray*}
So \kak{arai} follows for $f\in C_0^\infty(\RR)$.
By a limiting argument on $f$
and the fact that $T$ is closed,
\kak{arai} follows for $f\in C^1(\RR)$ such that $f$ and $f'$ are bounded.
\qed
This proposition suggests that
{\it informally}
 $$Te^{-itg(H)}\phi=e^{-itg(H)} T\phi+t g'(H)e^{-itg(H)} \phi$$
and then
$Tg'(H)^{-1} e^{-itg(H)}\phi=
e^{-itg(H)}(Tg'(H)^{-1}+t) \phi$.
Symmetrizing
$Tg'(H)^{-1}$, we expect
that a strong time operator associated with $g(H)$ will be given by
\eq{es}
T_{g(H)}=\half (g'(H)^{-1}T+Tg'(H)^{-1}).
\en
In order to establish \kak{es}, the remaining problem
is to check the  domain argument and to extend Proposition
\ref{l1} for unbounded $f$ and $f'$.

By the definition of $g$,
for $\la\in \RR\setminus Z$,
there exists the derivative $dg(\la)/d\la=g'(\la)$
and $g'(\la)^{-1}<\infty$.
Let
\eq{def1}
\tilde g'(\la)=
\lkk\begin{array}{ll}g'(\la),&\la\not\in Z,\\
0,&\la\in Z
\end{array}
\right.
\en
and
define
\eq{def2}
g'(H)=\tilde g'(H).
\en
Equivalently 
\eq{def4}
g'(H)=\int_{\rm spec(H)\setminus Z} g'(\la) dE_\la^H.
\en
In what follows we denote $g'(\la)$ for $\tilde g'(\la)$ without confusion may arise.
Since the Lebesgue measure of $Z$ is zero and $H$ has purely absolutely continuous spectrum,
we see that
$${\rm dim}\  {\rm ker} g'(H)=0.$$
Thus $g'(H)^{-1}$ is well defined.
\bl{l2}
It follows that
\begin{description}
\item[(1)]
 $T:X_n^\dt \rightarrow X_{n-1}^\hhh$ for
$1 \leq n \leq \infty$.
\item[(2)]
$\g :
\lkk
\begin{array}{ll}
X_n^{\ms D} \rightarrow X_1^{\ms D},& 1 \leq n \leq \infty,\\
X_0^{\ms D} \rightarrow X_0^{\ms D},& n=0,
\end{array}
\right.$
for any $\ms D\subset \hhh$.
\end{description}
\el
\proof
Let $\Phi=\rho(H)\phi\in X_n^\dt$.
By Proposition \ref{l1},
 $\Phi\in \dm(T)$ and we have $T\Phi=i\rho'(H)\phi+\rho(H)T\phi$. Then (1) follows.
It is clear that $D(g'(H)^{-1})\ni \Phi=\rho(H)\phi$
and
$g'(H)^{-1}\Phi=(g'(H)^{-1}\rho(H))\phi$.
Note that
$\rho/g'\in
C_0^1(\RR\setminus Z)$ for $\rho\in C_0^n(\RR\setminus Z)$ with $n\geq1$,
and $\rho/g'\in C_0(\RR\setminus Z)$ for
$\rho\in C_0(\RR\setminus Z)$.
Then (2) follows.
\qed
Define
the symmetric operator $\widetilde D$ by
\eq{4}
\widetilde D=\left.
\half (\g T + T \g) \right \lceil_{X_1^\dt}.
\en
$\widetilde D$ is well defined by Lemma \ref{l2}. Actually $\widetilde D: X_1^{D(T)}\rightarrow X_0^\ms H$.
Since the domain of the adjoint of $\widetilde D$ includes the dense subspace $X_1^\dt$,
 $\widetilde D$ is closable.
We define
 \eq{115}
 D=
\half \ov{\left.(\g T + T \g) \right \lceil_{X_1^\dt}}.
\en
The main theorem is as follows.
\bt{main1}
Suppose Assumptions
\ref{a1} and \ref{a2}.
Then $(g(H), D)$ obeys the weak Weyl relation.
\et
\begin{example}
Examples of strong time operators are as follows:
\begin{description}
\item[(1)] Let $g$ be a polynomial. Then
$Z=\{\la\in \RR|g'(\la)=0\}$ and
a strong time operator associated with $g(H)$
is $$\half\ov{\lk g'(H)^{-1}T+Tg'(H)^{-1}\rk
\lceil_{X_1^{D(T)}}}$$
\item[(2)] Let $g(\la)=\log|\la|$.
Then $Z=\{0\}$ and a strong time operator associated with $\log|H|$ is $$\half\ov{ (HT+TH)\lceil_{X_1^{\dm(T)}}}.$$
This strong time operator is derived in  {\rm \cite{am08}}.
\item[(3)]
Let $(H,T)=(P,Q)$ and $g(\la)=\sqrt{\la^2+m^2}$, $m\geq 0$. Then $Z=\lkk
\begin{array}{ll} \emptyset, & m>0\\ \{0\},& m=0
\end{array}
\right..$
A~strong time operator
 associated with $H(P)=\sqrt{P^2+m^2}$ is
 $$    \half\ov{(H(P)P^{-1}Q+QP^{-1}H(P))
    \lceil_{\dm(X_1^{\dm(Q)})}}.$$
        $H(P)$ is a semi-relativistic Schr\"odinger operator.

\item[(4)]
(3)
can be generalized to fractional Schr\"odinger operators.
Let
$\alpha\in \RR\setminus\{0\}$.
Define $H_\alpha(P)$ by
 $H_\alpha(P)=(P^2+m^2)^{\alpha/2}$.
     A strong time operator associated with $H_\alpha(P)$ is given by
$$\frac{1}{2\alpha}
\ov{
\lk
(P^2+m^2)P^{-1}H_\alpha(P)^{-1}Q+QH_\alpha(P)^{-1}P^{-1}(P^2+m^2)
\rk\lceil_{\dm(X_1^{\dm(Q)})}}.$$
\end{description}
\end{example}

\section{Proof of Theorem \ref{main1}}
In order to prove Theorem \ref{main1} we prepare two 
lemmas,
where
it is proven that
the weak Weyl relation holds for the pair $(g(H), \tilde D)$ but on $X_1^{D(T)}$.
\bl{l3}
Let $\Phi\in X_1^\dt$. Then
\begin{description}
\item[(1)]
$\Phi\in \dm(\gbb)$ and
$\gbb\Phi\in \dm(T)$;
 \item[(2)]
  $\gbb \e \Phi\in \dm(T)$;
  \item[(3)]
    $\e\Phi\in \dm(T)$ and $T\e \Phi\in \dm(\gbb)$.
 \item[(4)]
  $\e T\Phi\in \dm(\gbb)$;
\end{description}
\el
\proof
Throughout the proof we set $\Phi=\rho(H)\phi\in X_1^{D(T)}$ with some $\rho\in C_0^1(\RR\setminus Z)$ and $\phi\in D(T)$.
Note that $g\in C^2(\RR\setminus K)$.

(1)
Since
$
\rho/g '\in C_0^1(\RR\setminus Z)
$,
 $\gbb \Phi=(\gbb \rho(H))\phi
  \in \dm(T)$ follows from
Proposition \ref{l1}.

(2) Since
  $e^{-itg }\rho/ g ' \in C_0^1(\RR\setminus Z)$,
     $\e \gbb \Phi\in \dm(T)$ also follows from Proposition~\ref{l1}.

(3) Since $\xi=e^{-itg }\rho\in C_0^1(\RR\setminus Z)$
and its derivative is bounded, $\e\Phi\in \dm(T)$  and
$$T\e \Phi=T\xi (H)\phi=\xi(H)T\phi+i\xi'(H)\phi$$ follows from Proposition \ref{l1}.
Here $\xi'=-itg' e^{-itg}\rho+e^{-itg}\rho' \in C_0(\RR\setminus Z)$.
From this we have
$T\e \Phi\in \dm(\gbb)$.

(4) Since
$T\Phi=T\rho(H)\phi=i\rho'(H)\phi+\rho(H)T\phi$ and then
$$\e T\Phi=i\e \rho'(H)\phi+\e \rho(H)T\phi,$$
we have $\e T\Phi\in \dm(\gbb)$.
\qed

\bl{l66}
Let $\Phi\in X_1^\dt$. Then
\eq{ast}
\tilde D  \e \Phi=\e (\tilde D  +t)\Phi.
\en
\el
\proof
Let $\Phi=\rho(H)\phi\in X_1^{D(T)}$ with some $\rho\in C_0^1(\RR\setminus Z)$ and $\phi\in D(T)$.
 We divide the proof into three steps.

(Step 1)
It holds that 
\eq{chi}
T\gbb \e \Phi=\e (T \gbb +t) \Phi.
\en
{\it Proof:}
Let $\xi=e^{-itg}\rho\in C_0^1(\RR\setminus Z)$. As was seen in the proof of (3) of Lemma \ref{l3},
both $\xi$ and $\xi'$ are bounded and
\eq{im8}
T\e \Phi=T\xi(H)\phi=\xi(H)T\phi+i\xi'(H)\phi.
\en
Here
\eq{im9}
\xi'(H)\phi=-itg'(H)\e \rho(H)\phi+\e \rho'(H)\phi.
\en
Then \kak{im8} and \kak{im9} yield that
\eq{im10}
T\e\Phi=tg'(H)\e \rho(H)\phi+\e(\rho(H)T\phi+i\rho'(H)\phi).
\en
Note that $T\Phi=T\rho(H)\phi=\rho(H)T\phi+i\rho'(H)\phi$.
Then we have
\eq{5}
T\e \Phi=\e (T+t g'(H)) \Phi.
\en
Since we have already shown in (1) and (2) of Lemma \ref{l3}
that
$\Phi\in \dm(\gbb)$ and
$\gbb \Phi\in \dm(\e T)\cap
\dm(T\e)$,
we can substitute $\gbb\Phi$ for $\Phi$ in \kak{5}.
Then \kak{chi} follows.

(Step2)
It holds that 
\eq{6}
\gbb T \e \Phi=\e (\gbb T +t )\Phi.
\en
{\it Proof:}
Let
$\Psi\in X_1^\dt$.
\kak{chi} implies that
\eq{27}
(\Phi, T \gbb \e \Psi-\e T \gbb \Psi)=t(\Phi, \e\Psi).
\en
By (3) and (4) of Lemma \ref{l3}, we can take the adjoint of
both sides of \kak{27}.
Then \kak{6} follows if we transform $t$ to $-t$.

(Step3) Combining \kak{chi} and \kak{6}, we have \kak{ast}. 
\qed

{\it Proof of Theorem \ref{main1}:}

Let $\Phi\in \dm(D)$. There exists
$\Phi_n\in X_1^\dt$ such that
$\Phi_n\rightarrow \Phi$ and
$D\Phi_n\rightarrow D\Phi$ as $n\rightarrow \infty$ strongly.
By Lemma \ref{l66},
for each $\Phi_n$,
$De^{-itg(H)}\Phi_n=e^{-itg(H)}(D+t)\Phi_n$ holds.
Since $D$ is closed, the theorem follows by
 a limiting argument.
\qed

\noindent {\bf Acknowledgments:}
We thank A. Arai for helpful
comments and careful reading of the first manuscript.
We also thank unknown referee for useful comments.


{\footnotesize
\begin{thebibliography}{99}

\bibitem[Ara05]{a05}
A. Arai, Generalized weak Weyl relation and decay of quantum dynamics, {\it Rev. Math. Phys.} {\bf 17} (2005), 1071--1109.


\bibitem[Ara08]{a08}A. Arai,
On the uniqueness of weak Weyl representations of
the canonical commutation relation, to be published in
{\it Lett. Math. Phys.}



\bibitem[Ara08-b]{a08b}
A. Arai,
Necessary and sufficient conditions for a Hamiltonian with discrete eigenvalues to have time operators,
 mp-arc 08-154, preprint 2008.




 \bibitem[AM08-a]{am08}
A. Arai and Y. Matsuzawa,
Construction of a Weyl representation from a weak Weyl representation of the canonical commutation
relation, {\it Lett. Math. Phys.} {\bf 83} (2008), 201-211.
\bibitem[AM08-b]{am081} A. Arai and Y. Matsuzawa,
    Time operators of a Hamiltonian with purely discrete spectrum,
        to be published in
{\it Rev. Math. Phys.}



\bibitem[Gal02]{gal02}
 E. A. Galapon, Self-adjoint time operator is the rule for discrete semi-bounded Hamiltonians,
{\it Proc. R. Soc. Lond.} {\bf  A 458} (2002), 2671--2689.

\bibitem[Gal04]{gal04}
 E. A. Galapon, R. F. Caballar and R. T. Bahague Jr, Confined quantum time of
arrivals, {\it Phys. Rev. Lett.}{\bf  93} (2004), 180406.


\bibitem[Dor84]{dor84}G. Dorfmeister and J. Dorfmeister,  Classification of certain pairs of operators $(P,Q)$ satisfying $[P,Q]=-i{\rm Id}$,
    {\it J. Funct. Anal.} {\bf 57} (1984), 301--328.
\bibitem[Fuj80]{fuj80}
I. Fujiwara, Rational construction and physical signification of the quantum time operator, {\it Prog. Theor. Phys.} {\bf 64} (1980), 18--27.
\bibitem[FWY80]{fwy80}
I. Fujiwara, K. Wakita and H. Yoro,
Explicit construction of time-energy uncertainty relationship in quantum mechanics, {\it Prog. Theor. Phys.} {\bf 64} (1980), 363--379.


\bibitem[GYS81-1]{gys81-1} T. Goto, K. Yamaguchi and N. Sudo, On the time opertor in quantum mechanics,
{\it Prog. Theor. Phys.} {\bf 66} (1981), 1525--1538.

\bibitem[GYS81-2]{gys81-2} T. Goto, K. Yamaguchi and N. Sudo, On the time opertor in quantum mechanics. II,
{\it Prog. Theor. Phys.} {\bf 66} (1981), 1915--1925.



\bibitem[KA94]{dk94}
D. H. Kobe and V. C. Aguilera-Navarro,
Derivation of the energy-time uncertainty relation.
{\it Phys. Rev.} {\bf  A 50} (1994), 933 - 938.
\bibitem[LLH96]{llh90}H. R. Lewis, W. E. Laurence and J. D. Harris, Quantum action-angle variables for the harmonic oscillator, {\it Phys. Rev. Lett.} {\bf 26} (1996), 5157-5159.


\bibitem[Miy01]{m01}
M. Miyamoto, A generalised Weyl relation approach to the  time operator and its connection to the survival probability,
{\it J. Math. Phys.} {\rm 42} (2001), 1038--1052.

\bibitem[Ros69]{ros69}D. M. Rosenbaum,
Super Hilbert space and the quamntum-mechanical time operators,
{\it J. Math. Phys.}{\bf 19} (1969), 1127--1144.

\end{thebibliography}
}
\end{document}